\documentclass[prl,twocolumn,letterpaper,amsmath,amssymb,amsfonts,superscriptaddress,showpacs]{revtex4}

\usepackage{graphicx}
\usepackage{epstopdf}
\usepackage{amsmath}
\usepackage{bbm}

\newcommand{\be}{\begin{equation}}
\newcommand{\ee}{\end{equation}}
\newcommand{\ba}{\begin{align}}
\newcommand{\ea}{\end{align}}

\newcommand{\ket}[1]{\ensuremath{| #1 \rangle}}
\newcommand{\prj}[1]{\ensuremath{| #1 \rangle \langle #1 |}}

\begin{document}

\title{Entanglement dynamics in open two-qubit systems via diffusive quantum trajectories}

\author{Carlos Viviescas}
\affiliation{Departamento de F\'{\i}sica, Universidad Nacional de Colombia,
  Carrera 30 No.45-03 Edificio 404, Bogot\'a D.C., Colombia}
\author{Ivonne Guevara}
\affiliation{Departamento de F\'{\i}sica, Universidad Nacional de Colombia,
  Carrera 30 No.45-03 Edificio 404, Bogot\'a D.C., Colombia}
\author{Andr\'e R.R. Carvalho}
\affiliation{Australian Centre for Quantum-Atom Optics, Department of Quantum Science, Research School of Physics and Engineering, The Australian National University, Canberra, ACT 0200, Australia}
\author{Marc Busse}
\affiliation{Institut f\"ur Theoretische Physik, Ludwig-Maximilians-Universit\"at M\"unchen, Theresienstrasse 33, D-80033 M\"unchen, Germany}
\author{Andreas Buchleitner}
\affiliation{Physikalisches Institut der Albert-Ludwigs-Universit\"at Freiburg, Hermann-Herder-Strasse 3, D-79104 Freiburg, Germany}

\date{\today}

\begin{abstract}

We use quantum diffusive trajectories to prove that the time evolution of two-qubit entanglement under spontaneous emission can be fully characterized by \emph{optimal}  continuous monitoring. We analytically determine this optimal unraveling and derive a deterministic evolution equation for the system's concurrence. Furthermore, we propose an experiment to monitor the entanglement dynamics in bipartite two-level systems and to determine the disentanglement time from a single trajectory.

\end{abstract}
\maketitle

The fragility of entanglement with respect to decoherence persists as one of the major obstacles faced by most implementations of quantum information protocols. The unavoidable coupling of quantum systems to their surroundings is responsible for the deterioration of this most important quantum
information resource, in ways that still escape quantification, and of which we have only a limited understanding. A complete account of this degradation process, even for the most 
elementary systems, is hampered, to a large extent, by the optimization problem inherent to the characterization of mixed state entanglement \cite{Wern89,Uhlm00,Mint05}. 
This is one of the major current challenges for theory and experiment. 

Recently, new light was shed on open systems entanglement dynamics with the use of quantum trajectories \cite{Nha04,Carv07}. Based on numerical results, it was conjectured \cite{Carv07} that an \emph{optimal} continuous measurement prescription generates the -- in the above sense -- optimal pure state decomposition of the system state, such that the mean entanglement yields the correct mixed state entanglement, at all times. Such optimal unravelings may help to overcome the mentioned experimental and 
theoretical shortcomings, since they allow for a restriction of the optimization space, and define 
continuous monitoring prescriptions for the experimental measurement of entanglement.

In the present paper we show that the use of diffusive quantum trajectories allows for an analytical proof of the existence of optimal unravelings in the sense of \cite{Carv07}, for two qubits coupled to a zero temperature environment. Our results, however, go beyond this point, complementing the program proposed in \cite{Nha04,Carv07} in three further directions. First, we deduce a deterministic differential equation for the evolution of entanglement in open systems. To our knowledge, the first of its kind. Second, we show that a single quantum trajectory unambiguously determines the disentanglement time~\cite{Zycz01,Simon02,Dur04,Yu04,Carv04}, i.e. the time at which entanglement disappears completely. Finally, we propose an optical experimental realization of the optimal measurement strategy, 
for the actual monitoring of entanglement under incoherent dynamics. 

Consider a quantum system interacting with its environment. Initially prepared in a pure state $\ket{\Psi_0}$, it will evolve into a mixed state $\rho_t$. The temporal change 
of the entanglement inscribed into the system state can be inferred by evaluation of some 
entanglement measure $\mathcal{M}(\rho_t)$ at all times $t$, with ~\cite{Uhlm00}\begin{equation}
\label{eq:Mmix}
\mathcal{M}(\rho)=\inf_{\{p_i,\ket{\Psi_i}\}} \sum_i p_i \, \mathcal{M}(\Psi_i)\;,
\end{equation}
and $\mathcal{M}(\Psi)$ one of the known entanglement measures for pure states~\cite{Mint05}. The \emph{infimum} is understood over all possible decompositions of $\rho$ into
pure states, $\rho=\sum_i p_i \prj{\Psi_i}$, with positive $p_i \le 1$ and $\sum_i
p_i=1$. This optimization 
turns into a formidable numerical problem for higher dimensional or multipartite systems. The drawbacks, however, are not only computational. Even in those
cases where \eqref{eq:Mmix} can be analytically evaluated 
\cite{Woot98}, e.g., for a system composed of two qubits, no clear physical interpretation of this optimal decomposition exists such as to guide our intuition 
for the physics of mixed state entanglement. We believe that our present 
quantum trajectory approach indicates a possible strategy to 
overcome both these computational and conceptual obstacles.

We analyse open quantum systems described by a Lindblad master equation~\cite{Lind76}
\begin{equation}
\label{eq:LME} 
\dot\rho= -\frac{i}{\hbar} [H,\rho] + \sum_{k=1}^{N} \mathcal{L}_k \rho.  
\end{equation}
Here, the first term on the right hand side accounts for the unitary dynamics generated by the system Hamiltonian $H$, while the superoperators $\mathcal{L}_k$ describe the effects of the environment on the system. Their action on an arbitrary state $\rho$ is
\begin{equation*} 
  \mathcal{L}_k \rho = J_k \rho J_k^{\dagger} -
  \frac{1}{2} (J_k^{\dagger}J_k \rho + \rho J_k^{\dagger}J_k)\,,
\end{equation*}
and the operators $J_k$ -- which we will lump together as a vector $ \mathbf{J}$ hereafter -- are determined 
by the specific coupling between system and environment.

Instead of attempting a direct solution of \eqref{eq:LME} for the mixed state $\rho$, we follow an alternative path in the spirit of \cite{Carv07}: A continuous \emph{monitoring} of the environment yields information on the system, and induces a stochastic state evolution \cite{Carm93,Perc98,Wise01} conditioned on the measurement record. The unconditional state $\rho$ is recovered upon averaging the conditioned state $\ket{\Psi_c}$ over many independent realizations of the quantum trajectories at some given time $t$. Hence, the master equation is \emph{unraveled} into quantum trajectories, with different unravelings arising from different ways to measure the environment \cite{Carm93}. It is this last feature which is at the core of our work. In what follows, we rigorously show that an optimal measurement scheme can be found that generates physically motivated decompositions of $\rho$ into pure states, with minimal entanglement average equal to the system entanglement as defined by Eq.~\eqref{eq:Mmix}.

Our focus is on diffusive unravelings, yielding a continuous evolution for $\ket{\Psi_c}$. The corresponding It\^{o} stochastic equation for the state increment is 
\cite{Rigo97,Wise01,Wise05,Perc98}
\begin{equation}
\label{eq:SSE}
\begin{split}
  d\ket{\Psi_c} =& \left[-\frac{i}{\hbar}H -\frac{1}{2} \left(\mathbf{J}^{\dagger} \mathbf{J}
      + \langle \mathbf{J}^{\dagger}\rangle_{c} \langle \mathbf{J}
      \rangle_{c} -2 \langle \mathbf{J}^{\dagger}\rangle_{c}
      \mathbf{J} \right)\right] \ket{\Psi_c} \,dt \\
  &+ d\boldsymbol{\xi}^{\dagger}(t) (\mathbf{J} - \langle \mathbf{J}
  \rangle_{c} )\ket{\Psi_c} \,,
\end{split}
\end{equation}
where all expectation values are taken with respect to the conditional state $\ket{\Psi_c}$. The first term describes the
deterministic dynamics of the system, while the second, proportional to the vector $d\boldsymbol{\xi}=(d\xi_1,\dots,d\xi_N)^{\mathsf{T}}$ composed of 
infinitesimal complex Wiener increments \cite{Gard85}, encompasses the stochastic nature of the evolution.
The stochastic process $d\boldsymbol{\xi}$ has vanishing  ensemble average,
$E[d\boldsymbol{\xi}]=0$,  
with correlations 
\begin{equation}
\label{eq:dxi}
d\boldsymbol{\xi}d\boldsymbol{\xi}^{\dagger}= \mathbb{I}\,dt\,, \quad
d\boldsymbol{\xi}d\boldsymbol{\xi}^{\mathsf{T}}= u\, dt\; ,
\end{equation}
where $\mathbb{I}$ is the identity matrix and $u$ a complex symmetric matrix subject to the condition $\|u\|_2 \le 1$ for the matrix two-norm \footnote{The matrix two-norm $\|A\|_2$ of a square complex matrix $A$ is defined as the squared root of the maximum eigenvalue of $A^{\dagger}A$. }.

Associated with the quantum diffusive process \eqref{eq:SSE} is the  
measurement record 
which conditions the evolution of $\ket{\Psi_c}$,
and which can be written as a vector of complex functions  
\begin{equation}
\label{eq:current}
  \mathbf{Y}^{\mathsf{T}}\, dt= \langle \mathbf{J}^{\dagger} u  +
    \mathbf{J}^{\mathsf{T}} \rangle_{c} \, dt + d\boldsymbol{\xi}^{{\mathsf{T}}}\,,
\end{equation} 
where each vector component represents a specific detection event. The conditional change in $\ket{\Psi_c}$ is obtained after using this experimental record to substitute the noise term $d\boldsymbol{\xi}$ in \eqref{eq:SSE}.

Notice that with the parametrization \eqref{eq:dxi}, a diffusive unraveling defined by (\ref{eq:SSE}) is completely specified once the correlation matrix $u$ is given. Different unravelings, even though equivalent and leading to the same unconditional solution of the master equation \eqref{eq:LME}, generate ensembles of trajectories with distinct  statistical properties. In particular, the {\em optimal} unraveling we are seeking must generate, at each time step, an optimal pure state decomposition of $\rho_t$ which minimizes the average entanglement $E[\mathcal{M}]\,$. Consequently, the necessary condition for an optimal unraveling is that it has to minimize the average differential  $E[d\mathcal{M}]\,$ of the entanglement measure. This is achieved by optimization of the parameters which define $u$.

Let us illustrate the ideas above by a specific example. Consider two qubits each coupled to a private bath, which we assume to induce decoherence by spontaneous emission (zero temperature reservoir). The corresponding Lindblad operators are $\mathbf{J}=(\sqrt{\gamma}\sigma_{-}^{(1)}\otimes \openone, \sqrt{\gamma}\openone\otimes\sigma_{-}^{(2)})^{\mathsf{T}}$, where $\sigma_{-}^{(i)}$ is the deexcitation operator of the $i$th qubit and $\gamma$ the spontaneous decay rate, which we assumed equal for both qubits. 

We will employ concurrence $c$ \cite{Woot98} to quantify the system entanglement so far denoted by $\mathcal{M}$. Analogous results are obtained for other measures such as, e.g., entanglement of formation \cite{Woot98,Mint05}.
The concurrence of a general pure state $\ket{\Psi} = \psi_{00}\ket{00} + \psi_{01}\ket{01} + \psi_{10}\ket{10} + \psi_{11}\ket{11}$
is  
$c(\Psi)\equiv |\langle \Psi^{*} |\sigma_{y} \otimes \sigma_{y} |\Psi \rangle | = 2|\psi_{10}\psi_{01} - \psi_{00}\psi_{11}|$ \cite{Woot98}. 
Under the dynamics generated by 
Eq.~\eqref{eq:SSE}, the change 
of concurrence reads
\begin{equation}
  \label{eq:dC}
  \begin{split}
  dc_{\Psi} = & \left[-\gamma c_{\Psi} + 2\gamma
    \mathrm{Re}\left(\frac{\bar{c}_{\Psi}\psi_{11}^{*2}}{c_{\Psi}}u_{12}\right)
  \right] dt \\
  &-2 c_{\Psi} \mathrm{Re} (\langle \mathbf{J}^{\dagger} \rangle_{\Psi}d\boldsymbol{\xi})\,,
	\end{split}
\end{equation}
where 
$c_{\Psi}=c(\Psi)$, and $\bar{c}_{\Psi}= \langle \Psi^{*} |\sigma_{y} \otimes \sigma_{y} |\Psi \rangle $ 
\cite{Woot98}.

The optimal unraveling is now found by minimization of the ensemble average $E[dc_{\Psi}]$. This is accomplished by imposing that, for all trajectories and all times, the deterministic term in the first line of \eqref{eq:dC} be minimal. It is readily verified that this is achieved by setting
\begin{equation}
\label{eq:uop}
u^{\text{(opt)}} = \begin{pmatrix}
0 &  -e^{i\theta_{\mathrm{opt}}} \\
-e^{i\theta_{\mathrm{opt}}} & 0  
\end{pmatrix}\;,
\end{equation}
with $\theta_{\mathrm{opt}}=\mathrm{arg}
(\bar{c}_{\Psi}^{*}\psi_{11}^{2})$. 
This choice of parameters 
implies 
that the optimal unravelling, determined by the phase of $\Theta_{\Psi}\equiv\bar{c}_{\Psi}^{*}\psi_{11}^{2}$, 
is time independent: 
The equation of motion of $\Theta_{\Psi}$ in its It\^{o} form can be derived as 
\begin{equation*}
d\Theta_{\Psi}= \left[2 \gamma |\psi_{11}|^{4} u_{12} + \Theta_{\Psi} \mathrm{Re}( f_{\Psi}) \right] \, dt + \Theta_{\Psi} \mathrm{Re}(\mathbf{g}_{\Psi}^{\mathsf{T}}d\boldsymbol{\xi})\, ,
\end{equation*}
where $f_{\Psi}$ and $\mathbf{g}_{\Psi}$ are functions of $\ket{\Psi}$. Note that this equation depends on the unraveling through $u_{12}$. 
Upon substitution of $u^{\rm (opt)}$,  
it reduces to 
\begin{equation*}
d\Theta_{\Psi}=\Theta_{\Psi} \mathrm{Re} (\tilde{f}_{\Psi}\,dt + \mathbf{g}_{\Psi}^{\mathsf{T}}d\boldsymbol{\xi})\, ,
\end{equation*}
with $\tilde{f}_{\Psi}$ a new function of the state,  
and $d\Theta_{\Psi}$ thus has the same phase as $\Theta_{\Psi}$. Hence, along trajectories generated with the optimal unraveling, the phase of $\Theta_{\Psi}$ remains unchanged -- its value is fixed by the initial state. This feature has at least two significant consequences:
\begin{enumerate}
\item[(a)] First, the optimal decompositions for the unconditional state $\rho_t$ are singled out as ensembles of pure states  
which all exhibit the same phase of $\Theta_{\Psi}$, signaling the existence of a subtle relation between the temporal entanglement evolution and the 
characteristics of the optimal decompositions in Eq. \eqref{eq:Mmix}. 
\item[(b)] Second, 
it implies that the optimal unraveling is {\em time independent}, with 
\begin{equation*}
\theta_{\text{opt}}=\mathrm{arg} (\Theta_{\Psi_0}) = \mathrm{arg} (\bar{c}_{\Psi_0}^{*}\psi_{11}^{2}(0))
\end{equation*}
determined by the initial state $\ket{\Psi_0}$ only. This remarkable property, observed numerically in \cite{Carv07} and analytically confirmed here, is of utmost importance for the design of experimental setups to directly monitor the entanglement evolution (see below): the monitoring strategy associated with the optimal unraveling is settled once the initial state is chosen, and does not require any steering along with the system's temporal evolution.  
\end{enumerate} 

As a last step, we still need to show that the optimal unraveling defined above indeed coincides with the state's mixed state entanglement as defined in  \eqref{eq:Mmix}, since the set of physically realizable decompositions of $\rho$ is only a subset of all possible decompositions \cite{Wise01b}.
Substitution of $u^{(\text{opt})}_{12}$ into 
\eqref{eq:dC} and 
a subsequent ensemble average leads to  
the equation of motion  
\begin{equation}
\label{eq:dEC1}
 \frac{dE[c_{\Psi}]}{dt}= -\gamma E[c_{\Psi}] - 2\gamma E[|\psi_{11}|^{2}]\, .
\end{equation}
The evolution equation of $E[|\psi_{11}|^{2}]$ is unraveling independent and can be derived to read $dE[|\psi_{11}|^{2}]/dt = - 2\gamma E[|\psi_{11}|^{2}]$, which integrates to $E[|\psi_{11}|^{2}]=|\psi_{11}(0)|^{2}e^{-2\gamma t}$. Thus Eq.~\eqref{eq:dEC1} reduces to
\begin{equation}
\label{eq:dEC2}
\frac{dE[c_{\Psi}]}{dt}= -\gamma E[c_{\Psi}] - 2 \gamma |\psi_{11}(0)|^{2}e^{-2\gamma t}\,,
\end{equation}
a deterministic equation of motion for the average concurrence, with parameters that depend on $\ket{\Psi_0}$ only. The solution of this equation is
\begin{equation}
\label{eq:ct}
E[c_{\Psi}](t,\Psi_0)=e^{-\gamma t}\left[c(0,\Psi_0) - 2 |\psi_{11}(0)|^2 (1 - e^{-\gamma t})\right] \,,
\end{equation}
which coincides with the known result \cite{Carv07,Fran06} for the specific physical scenario under study here. 

\begin{figure}[t]
\begin{center}
\includegraphics[scale=.75]{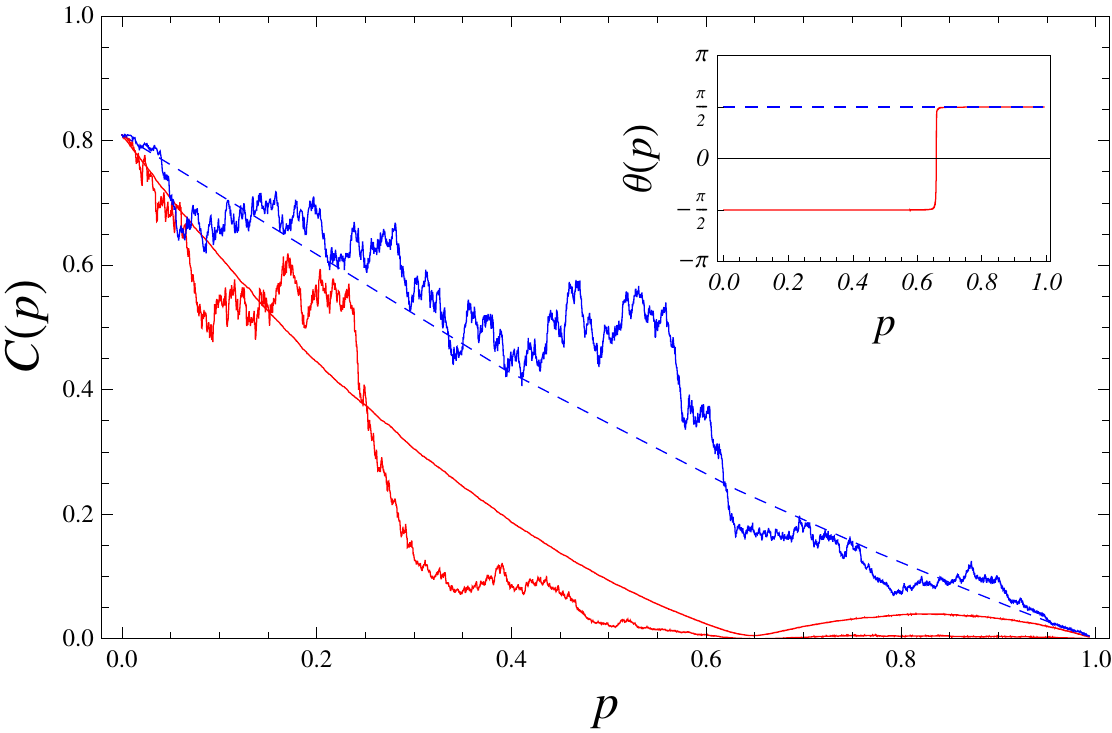}
\caption{Time evolution of concurrence for initial states
  $\ket{\Psi_0}=(i\sqrt{5}\ket{00} - \ket{01} + i \ket{10} +
  \ket{11})/\sqrt{8}$ (dashed line) and $\ket{\Psi_0}=(\ket{00} - \ket{01} + i
  \ket{10} + i\sqrt{5} \ket{11})/\sqrt{8}$ (solid line) under incoherent coupling to a zero temperature environment (time is rescaled as $p=1-e^{-\gamma t}$). Both states have the same initial concurrence $c(\Psi_0)=0.809$. The average concurrence evaluated using 500 trajectories with an optimal unraveling faithfully reproduces the exact result [cf.\ Eq.~\eqref{eq:ct}]. Irregular lines represent the pure stateconcurrence along individual quantum trajectories for each initial state. The inset shows the time evolution of $\theta=\mathrm{arg}(\bar{c}_{\Psi}^{*}\psi_{11}^{2})$ along these trajectories.  
}
\label{fig:QT}
\end{center}
\end{figure}
In addition to this coincidence between the average entanglement generated by optimal unraveling and mixed state entanglement according to \eqref{eq:Mmix}, even {\em single} quantum state trajectories carry relevant information on the disentanglement process induced by the environment coupling: As evident from Eq.~\eqref{eq:ct}, entangled initial states with $c(0,\Psi_0)<2|\psi_{11}(0)|^2$ 
turn separable at finite times $t_s<\infty$, while they only asymptotically reach separability otherwise \cite{Yu04,Carv07,Fran06} (see Fig.~\ref{fig:QT}). Since the concurrence of a two-qubit state is given by $c(\rho)=\mathrm{max}\{0, \Lambda(\rho) \}$ \cite{Woot98}, where $\Lambda(\rho)=\sqrt{\lambda_1} - \sum_{i=2}^{4} \sqrt{\lambda_i}$, and $\lambda_i$ are the eigenvalues of the matrix $\rho(\sigma_y \otimes \sigma_y)\rho^*(\sigma_y \otimes \sigma_y)$, with $\lambda_1 > \lambda_i$ for $i=2,3,4$, while $E[c_{\Psi}](t,\Psi_0)$ is obtained as a sum of positive numbers, the latter reproduces the \emph{modulus} of $\Lambda (t)\equiv \Lambda(\rho(t))$. Consequently, even $c_{\Psi}(t)$ as derived from an {\em individual} quantum trajectory vanishes when  $\Lambda (t)$ changes sign at $t=t_s$; this is associated with a jump of the phase $\theta(t) = \mathrm{arg} (\bar{c}_{\Psi}^{*}\psi_{11}^{2})$ by $\pi$. Thus, the disentanglement time $t_s$ is unambiguously encoded in each individual quantum state trajectory, as shown in Fig.~\ref{fig:QT}.

Let us finally describe an experimentally realizable setup for the direct monitoring of the open system entanglement evolution. For the optimal unraveling \eqref{eq:uop}, the measurement record \eqref{eq:current} has the explicit form 
\begin{equation}
\label{eq:optI}
\begin{split}
Y_1\, dt &= \sqrt{\gamma} \langle - e^{i\theta_{\mathrm{opt}}} \sigma_{+}^{(2)}
+ \sigma_{-}^{(1)} \rangle_c \, dt + d\xi_1 \,,\\
Y_2\, dt &= \sqrt{\gamma} \langle - e^{i\theta_{\mathrm{opt}}} \sigma_{+}^{(1)}
+ \sigma_{-}^{(2)} \rangle_c \, dt + d\xi_2 \,,
\end{split}
\end{equation}
where, for notational simplicity, we 
let the operators' superscripts specify the subspace on which they act. 
\begin{figure}[t]
\begin{center}
\includegraphics[scale=.4]{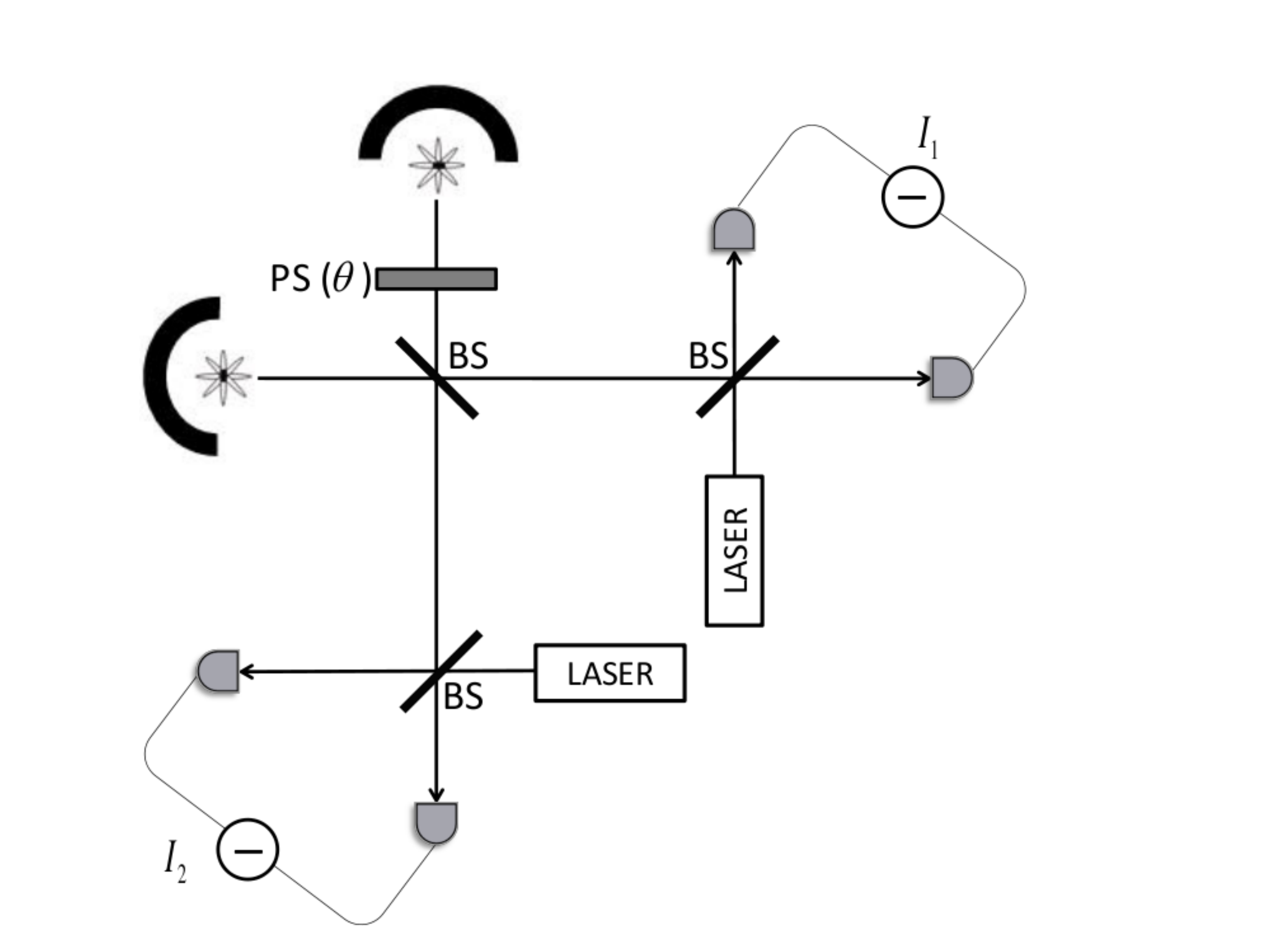}
\caption{Experimental setup for direct monitoring of two-qubit entanglement evolution under incoherent dynamics. One of the photons emitted by the two qubits passes through a phase shifter (PS) with $\theta=-\theta_{\text{opt}}+\pi/2$ before combining with the other photon in a 50-50 beam splitter (BS). Each of the outgoing modes is then subject to homodyne detection with a local oscillator.}
\label{fig:Exp}
\end{center}
\end{figure}
Such measurement can be realized in optical experiments:  Photons emitted by qubits $1$ and $2$, corresponding to $\sqrt{\gamma} \sigma_{-}^{(1)}$ and $\sqrt{\gamma} \sigma_{-}^{(2)}$, respectively, enter the detection scheme shown in Fig.~\ref{fig:Exp}. The photon coming from the second qubit passes through a phase shifter (PS) with $\theta=-\theta_{\text{opt}}+\pi/2$ before impinging on a 50-50 beam splitter (BS), where it is combined with the photon from qubit one. Each of the outgoing modes is then subject to homodyne detection with a local oscillator with its phase set equal to $\theta_{\text{loc}}=-\pi/2$. At the detectors, this procedure yields the real homodyne photocurrents~\cite{Wise93} 
\begin{equation}
\begin{split}
  I_1\, dt &= \sqrt{\frac{\gamma}{2}} \langle - e^{i\theta_{\mathrm{opt}}}
  \sigma_{+}^{(2)} +  \sigma_{+}^{(1)} + \text{h.c.} \rangle_c \, dt + d\zeta_1 \,,\\
I_2\, dt &= -i \sqrt{\frac{\gamma}{2}} \langle e^{i\theta_{\mathrm{opt}}}
  \sigma_{+}^{(2)} + \sigma_{+}^{(1)} - \text{h.c.} \rangle_c \, dt + d\zeta_2\,.
\end{split}
\end{equation}
Here, $d\boldsymbol{\zeta_i}$ are real, independent increments, corresponding to the detectors' shot noise. The complex currents \eqref{eq:optI} are recovered from these real currents through $Y_1\,dt =(I_1\, dt -i I_2\, dt)/\sqrt{2}$ and $Y_2\,dt =- e^{i\theta_\text{opt}}(I_1\, dt + i I_2\, dt)/\sqrt{2}\,$. The conditional evolution of the state is reconstructed by substituting these,
with the help of \eqref{eq:current}, into \eqref{eq:SSE}. This then allows for an immediate evaluation of the pure state concurrence. The mixed state entanglement evolution of the system is recovered after averaging over the ensemble of entanglement records generated in this way. 

This experimental proposal, together with the time evolution equation \eqref{eq:dEC2} and the finite time disentanglement detection with a single trajectory, convey the strength of diffusive quantum trajectories as a complete and efficient method for the characterization of entanglement under incoherent dynamics. Extensions of these ideas to other type of environments, higher dimensional systems, and considerations of finite detection efficiencies for experimental realizations are relevant issues to be addressed in future work. 

We thank Lajos Di\'osi,  Marek Ku\'s, Hans Maassen, Marco Piani, Dominique Spehner, Juan D.~Urbina, Howard M. Wiseman and Karol {\.Z}yzckowski for interesting and stimulating discussions. Financial support by VolkswagenStiftung and DAAD/PROBRAL is gratefully acknowledged.

\end{document}